\documentclass[11pt,draftclsnofoot,peerreviewca,onecolumn,twoside]{IEEEtran}

\usepackage[cmex10]{amsmath}
\usepackage{amsfonts}	
\usepackage{cite}
\usepackage{amssymb}
\usepackage[T1]{fontenc}
\usepackage{graphicx}
\usepackage{color, subcaption}

\usepackage{tikz,pgfplots}
\usetikzlibrary{positioning,shapes,babel}
\usetikzlibrary{external}
\tikzsetexternalprefix{Graficos/} 
\tikzsetfigurename{Fig} 
\pgfplotsset{tick label style={font=\scriptsize},
	legend style = {font=\small},
	xlabel style ={font = \footnotesize},
	ylabel style ={font = \footnotesize},
	yticklabel style={
		/pgf/number format/fixed,
		/pgf/number format/fixed zerofill
	},		scaled ticks = false,
	legend style = {cells={anchor=west}},
	grid = both,
	every axis plot/.append style={line width=1pt}
}

\pgfplotsset{compat=newest}
\begin{document}
\title{A Time-domain Approach to the Design of Coupled-Resonator Microstrip Filters}

\author{Andr\'es~Altieri
	\thanks{A. Altieri is with Universidad de Buenos Aires and CSC-CONICET, Buenos Aires, Argentina (email:  aaltieri@fi.uba.ar, andres.altieri@conicet.gov.ar).}
	\thanks{This work was partially supported by the Grant PICT 2016-1925 and the International Cooperation Project CNRS-CONICET MOSIME.}
}

\maketitle

\begin{abstract}
Coupled-resonator microstrip filters are among the most versatile filter topologies. A known design approach uses full-wave electromagnetic simulations to determine the coupling coefficient between resonators as a function of their relative position. This could be done using time-domain simulations using a fast Fourier transform (FFT) to extract the couplings from the S parameters obtained from time-domain signals. However, this approach has a poor performance in terms of resolution and specially for weak couplings, leading to unreasonably long simulation times. To overcome this, we introduce a technique to obtain the couplings directly from time signals, without moving to the frequency domain. This procedure works for strong and weak couplings, with much shorter simulation times and a reduced simulation domain over the FFT approach. This technique is used to design coupled resonator filter efficiently from time domain simulations. We implement this procedure using the finite-difference time-domain framework using an open source solver and discuss our implementation. We show that its results  are very similar to previously published ones obtained from frequency domain simulations, even in the case of very weak couplings. Finally, we design and measure a filter to show the good performance of the proposed approach. 

\end{abstract}

\begin{IEEEkeywords}
	Bandpass filter (BPF), hairpin filter, resonator, coupling coefficient, microstrip filter design, finite-difference time-domain.
\end{IEEEkeywords}

\section{Introduction and Main Contributions}

Coupled resonator filters~\cite{Hong_microstripbook} are one of the most compact and efficient topologies for microwave bandpass filter designs of small and moderate bandwidths. For this reason, they have been an active area of research for the last decades. A usual synthesis approach is to start with the desired filter response and obtain the required coupling coefficients that yield  this response. These include the  inner coupling coefficients~\cite{Tyurnev2010} between the resonators which compose the filter,
and the coupling between the feeding circuitry and the resonators, characterized by the external quality factor. Then, the dimension of the coupled resonators and their separation is chosen such that the prescribed couplings are obtained. 
The most complex procedure in the design is the sizing of the filter, that is, determining the dimensions and separation of the resonators to achieve the required couplings. Although for some filter topologies there exist approximate closed-form expressions~\cite{Cristal_Frankel1972, Gysel_1974, MATTHAEI}, in recent years numerical approaches to extract couplings  using full-wave been developed~\cite{Hong96, Hong2000, Wu2006} which are valid for many topologies. These techniques permit a fast and accurate design of filters with minimal tune-up.

In Fig. \ref{fig:coupledresonators}a we show a representation of the circuit to extract the coupling between two resonators. One of the resonators is excited via a weakly-coupled input circuit. This resonator elicits a coupled response on the other
 one, and this response is measured at an output port, weakly coupled to the resonator. In Fig. \ref{fig:coupledresonators}b, an example of such a circuit can be seen for the case of two microstrip hairpin resonators. Neglecting long range couplings, the input and output feed lines are coupled each to its adjacent  resonator and the resonators are coupled to each other. The level of  coupling between adjacent elements increases as the separations $s$ and $\tilde{s}$ decrease. This circuit can be interpreted as two-port circuit like Fig. \ref{fig:coupledresonators}c. 
Assuming that both resonators are equal, to extract the coupling coefficient in the frequency domain, the $S_{2,1}$ parameter of the circuit is obtained. If the feed circuitry is weakly coupled to the resonators ($s \ll \tilde{s}$ in the example of Fig. \ref{fig:coupledresonators}b), two peaks would be observed in $|S_{2,1}|$ near the resonance frequency $f_0$ of each resonator. In Fig. \ref{fig:S21} a typical plot of the obtained $|S_{2,1}|$ with to peaks at frequencies $f_-$ and $f_+$. Then, the coupling coefficient between the resonators can be obtained as \cite{Hong_microstripbook,Hong2000}:
\begin{equation}
	K = \frac{f_{+}^2 - f_{-}^2}{f_{+}^2 + f_{-}^2}.\label{eq:K_eq}
\end{equation}
\begin{figure}
	\centering
	{\includegraphics[width=0.4\linewidth]{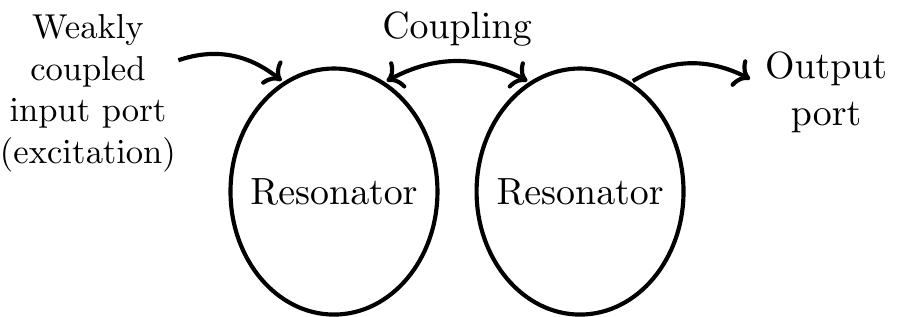}
		\subcaption{General setup for extracting the coupling coefficient.}
		
		\includegraphics[width=0.4\linewidth]{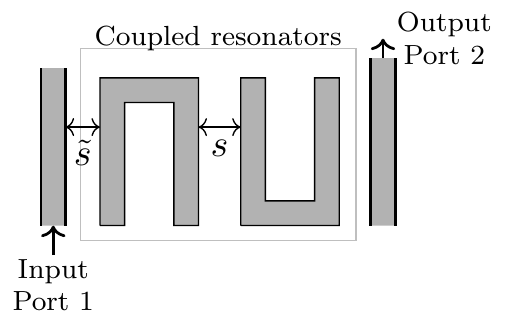}
		\subcaption{Example of implementation using hairpin resonators. }
		\includegraphics[width=0.3\linewidth]{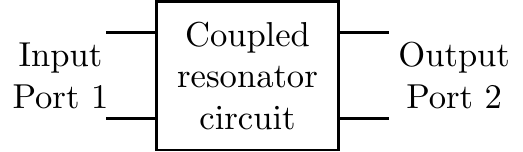}
		\subcaption{Interpretation as a 2 port-network of the above circuit.}}
		
	\caption{Circuit for extraction of the coupling coefficient between resonators.}
	\label{fig:coupledresonators}
\end{figure}
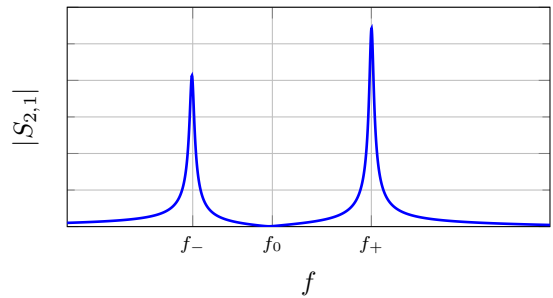
\begin{figure}[t!]
	\centering
	\begin{tikzpicture}
		\begin{axis}[scale=1,xlabel={$f$}, ylabel={$|S_{2,1}|$},height=4.5cm,width=8	cm,,legend pos = north east,
			ymin =0, ymax = 6000,xmin=2.8,xmax=4.8,minor x tick num=1,minor y tick num=1, grid =both, xtick={3.32,3.65,4.06},xticklabels={$f_-$, $f_0$, $f_+$}, ytick={0,2000,4000,6000},
			yticklabels={},
			legend style = {font = \scriptsize},]
			\addplot[blue,mark size =1.5pt,,line width=1pt]  table[x expr=\thisrowno{0}/1e9, y index=1] {Dat/MagZ_Ro4350_L1047_g30_w15.txt} ;
			\draw [dashed, thick] (axis cs: 2.113, 0) -- (axis cs: 2.113, 7000);
		\end{axis}
	\end{tikzpicture}
	\caption{Example of $S_{2,1}$ obtained near the resonance frequency of the circuit in Fig. \ref{fig:coupledresonators} near the resonance frequency $f_0$ of the resonators.  
	}\label{fig:S21}
\end{figure}
It is important to mention that, as the coupling decreases, the frequencies $f_-$ and $f_+$ become closer to each other. This means that the frequency resolution required to extract these couplings becomes increasingly large as the couplings decrease. Another important observation is that (\ref{eq:K_eq}) is very sensitive to small variations in $f_+$ and $f_-$, meaning that a small error in estimating them, will produce a large variation of the coupling coefficient $K$.

The procedure described in the previous paragraph can be used to experimentally measure the coupling coefficient of microstrip resonators. It can also be used to extract the coupling coefficient in frequency domain simulators. Depending on the simulator used, the downside of a frequency domain approach is that it requires the simulation of a large number of  discrete frequencies near the resonance frequency or relying on an effective interpolation of the spectrum. Implementing this procedure from time domain simulations, however, will introduce several complications. A straightforward approach  would be to obtain an estimate of the $S_{2,1}$ parameter using the Fast Fourier Transform (FFT) of the input and output signals. However, it is well known that the frequency resolution attainable when using periodogram-type estimators depends on the total number of samples available. In fact, the theoretical frequency resolution of the periodogram is $1/N$, where $N$ is the number of samples of the time signal. This resolution is poor compared to other approaches which achieve a much better resolution with a substantially smaller number of samples. Furthermore, it provides biased and inconsistent frequency estimations Finally, to estimate the transfer function we must perform the quotient of two inconsistent estimators, which further increases the final estimator variance~\cite{stoica2005spectral}.  This means that extracting the coupling coefficients from the time-domain simulations through a standard FFT approach is not practical or is sometimes unfeasible because it  requires very long simulation times and leads to poor estimates. 

In this paper we propose a technique to extract the coupling coefficient between microstrip resonators using time-domain simulations, which overcomes many of the difficulties mentioned in the previous paragraph. This technique will work for both strong couplings and weak couplings, where using the FFT is impractical.  In addition, in the strong coupling regime, where using the FFT approach is possible, the proposed technique will require much shorter simulation times and a reduced simulation domain. In principle it is possible to apply this technique to many resonator topologies, specially  those which are built from half-wavelength resonators, such as, hairpin resonators, square and rectangular open-loop resonators, among others (see \cite{Hong_microstripbook} for examples). Our tests with hairpin and square resonators show good agreement with other works and measurements. Finally, although we discuss the approach using equal resonators, the case for unequal resonators~\cite{Hong2000} can also be considered. 

In the proposed technique, one of the resonators is excited directly by injecting a current pulse at one of its edges. Then, the voltage time signal on the other resonator is measured by integrating the electric field between the ground plane and the resonator.
After processing the measured voltage, the resonant frequencies are extracted from the signal by applying a super-resolution technique~\cite{stoica2005spectral}, in our case ESPRIT~\cite{Kailath_ESPRIT}. This approach provides consistent frequency estimates with short simulation times. In addition, the absence of feeding circuitry will  shorten simulation times and the computational resources by reducing the size of the computation domain, and will mitigate the disturbance introduced by the feeding circuitry coupling. We implement the procedure using the finite-difference time-domain method (FDTD)~\cite{taf05} and compare the results obtained with previously published results from other authors. In addition, we also design and build a hairpin filter to validate the technique through measurements.
All our simulations are carried out using gprMax~\cite{WARREN2016163}, an open-source FDTD simulator,  which means that the results are easily reproducible. To this end, we also discuss details of our implementation. 

In the past, there have been works which use super-resolution techniques for the extraction of frequency domain information from time-domain simulations, but to the best of our knowledge they have not been applied to the analysis of resonant microstrip circuits or microstrip filter design. For example, in \cite{Ling1997, Chen2002}, the authors use super-resolution techniques on FDTD signals to extract frequency-dependent equivalent-circuit parameters of microstrip multiconductor transmission lines. Other approaches to extract frequency domain information from FDTD signals are for example \cite{Litva1994,Shaw2001}. In \cite{Litva1994}, the authors propose two approaches to extract models from FDTD signals: in one they treat the signals as random processes and construct spectral estimators based on autoregressive models. In the second one they construct a signal model based on a nonlinear fitting of the time-domain signals.  In \cite{Shaw2001}, the authors propose an iterative approach to generate an autoregressive moving-average model (ARMA) from a short time-domain simulation by minimizing the norm of the aproximation error. These models, however, are better suited for continuous spectra and are more sensitive to perturbations in the time signals. Super-resolution techniques, on the other hand, are better estimators for line spectra and they are more robust to noise and perturbation than time-fitting approaches because they are specifically tailored for this application~\cite{stoica2005spectral}.

%
%
%

The rest of the paper is organized as follows. In Section II we discuss the proposed approach to extract the couplings. We also review the procedure to extract the external quality factor, which is also required to design filters. In Section III we compare the proposed approach to previously published results by others authors and to measurements. Finally, in Section IV we present some closing remarks. 

\section{Procedure to design the resonator filters}

\subsection{Extraction of the coupling coefficient between resonators} \label{sec:extractionK}
\begin{figure}
	\centering
	\includegraphics[width=0.5\linewidth]{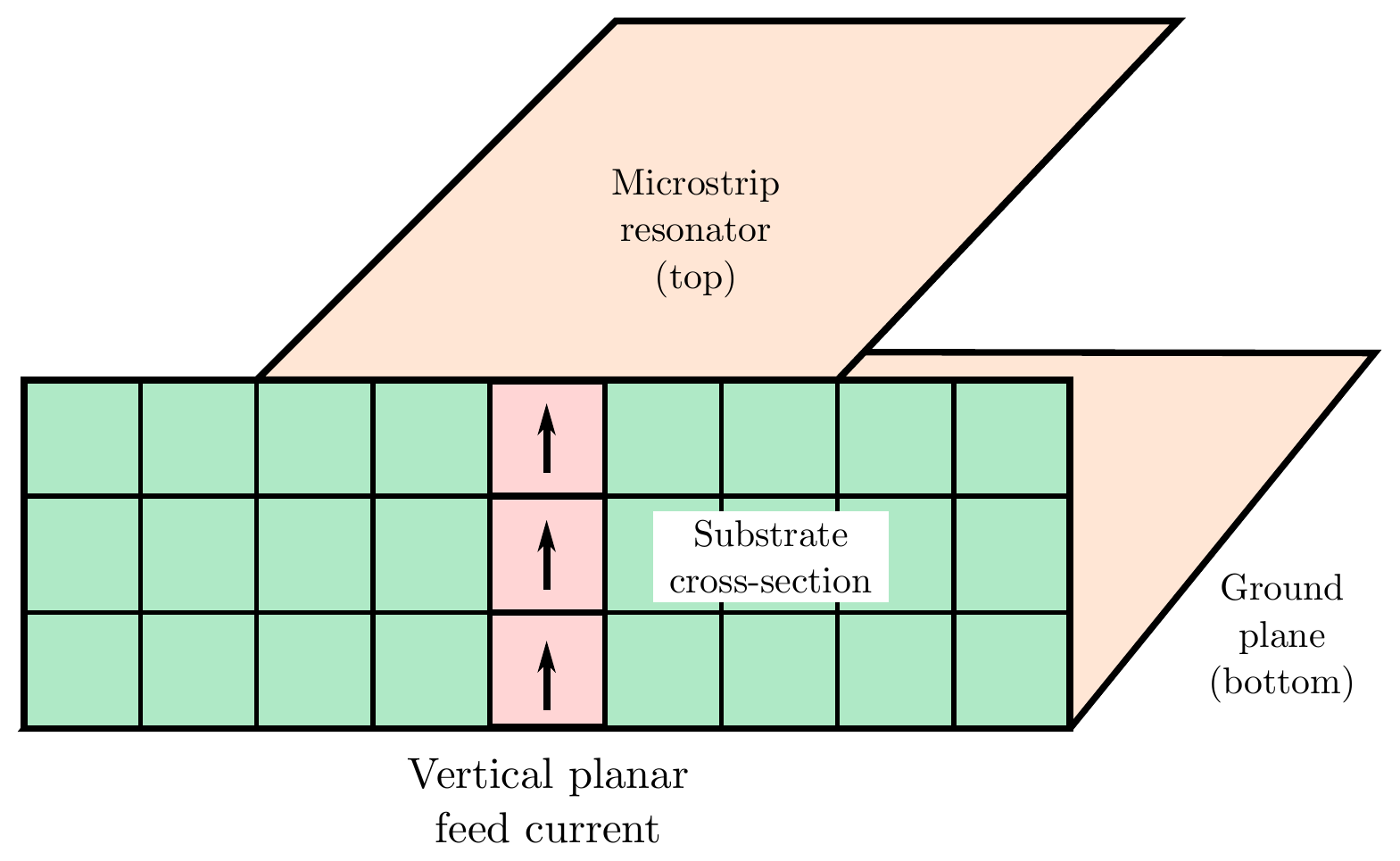}
	\caption{Microstrip resonator edge feed model. If the microstrip has an even number of cells, the feed current width is doubled to achieve simmetry.}
	\label{fig:feedmodelperspective}
\end{figure}

\begin{figure}\centering{
\begin{minipage}[]{0.45\columnwidth} \centering
	\includegraphics[width=.8\linewidth]{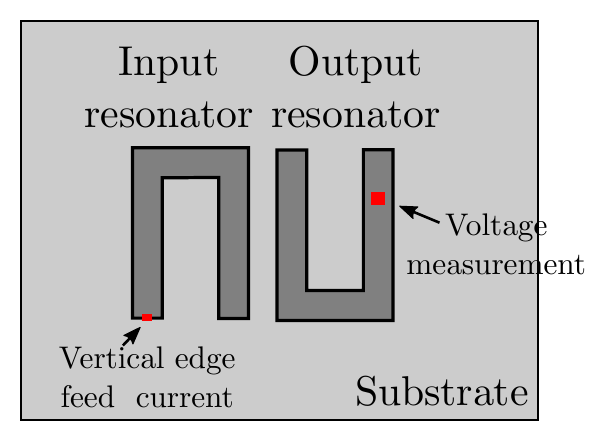}
	\subcaption{Hairpin resonator exhibiting mixed mode coupling. \phantom{xxxxx}.}
\end{minipage}
\begin{minipage}[]{0.45\columnwidth}\centering
	\includegraphics[width=.8\linewidth]{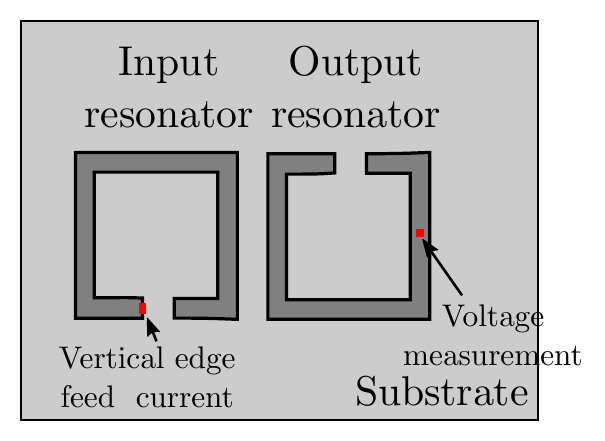}
	\subcaption{Square open loop resonator exhibiting mixed mode coupling.}
\end{minipage}}
\caption{Current excitation and voltage measurement example for two typical microstrip resonators using the proposed approach.} \label{fig:exc_example}
\end{figure}

In this section we describe the procedure to extract the coupling coefficient between two identical microstrip resonators which have a resonance frequency $f_0$. The coupling strength is controlled by how the resonators are laid out relative to each other. To elicit a resonant response, one of the resonators is excited injecting a current pulse between the ground plane and the microstrip resonator. This is done via a planar current source on one of the edges of the resonator. By directly injecting the excitation in the resonator, the feed lines are avoided, reducing the simulation domain size and avoiding the perturbation of the feed lines. In Fig. \ref{fig:feedmodelperspective} a representation of the current feed model can be seen for our FDTD implementation. We assume that the cell is perfectly aligned with the microstrip width, and the planar current is injected at the microstrip center. In gprMax this type of feed is achieved using the hertzian dipole source, which specifies a current density term at an electric field location. In our implementation, we use a Gaussian current pulse $p$ given by:
\begin{equation}
	p(t) = e^{-2 \pi f_p^2 \left(t-\frac{1}{f_p}\right)^2} \ \ \ (t>0), \label{eq:gaussian}
\end{equation}
where $f_p$ is a parameter which controls the time duration/bandwidth of the pulse. It is required that $f_p \gg f_0$ if the resonant behavior is to be excited. The excitation of the input resonator will couple with the output resonator generating an oscillatory response from where the two characteristic frequencies required to obtain the coupling coefficient can be extracted. To do this, the voltage is measured at the center of the other resonator by numerical integration of the vertical field between the ground plane and the microstrip. Since the excitation pulse will have an approximately flat frequency content in the range of interest, the resonant peaks in $|S_{2,1}|$ will be present in the raw voltage signal measured at the output resonator. Hence, it is not necessary to compute any S parameters, we can extract the frequencies directly from the voltage signal measured at the output resonator. In Fig. \ref{fig:exc_example} the location of the feed current and voltage  measurement can be seen for two types of  common resonators, namely, hairpin and square open-loop. 

\begin{figure}[t!]
\centering
\includegraphics[width=.8\linewidth]{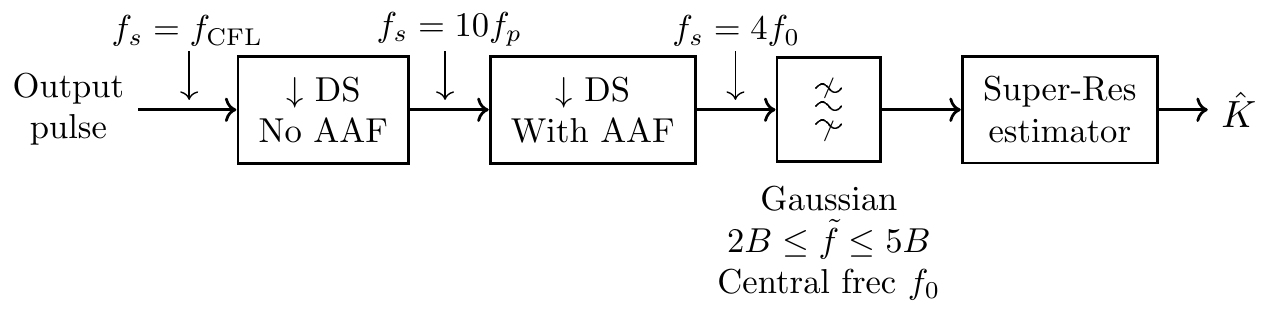}
\caption{Overview of the algorithm for extracting the coupling coefficient from the time-domain simulated output pulse. $f_s$ is the sampling frequency of the signal after each processing block. DS: downsampling block. AAF: anti-aliasing filter. The Gaussian bandpass filter has impulse response given by (\ref{eq:Gfilt}).}
\label{fig:kalgorithm}
\end{figure}
Using the voltage signal, the spectral peaks are extracted. In short, the time signal is downsampled and filtered in several stages, and then the resonance frequencies are extracted using a super-resolution technique, in our case, ESPRIT.  The filtering problem is not a trivial one. This is because one of the most important steps in the implementation of a super-resolution algorithm is the determination of the model order, that is, the number of frequencies that are to be estimated. There are different criteria involved~\cite{Stoica2004,Costa_Thakre_Roemer_Haardt_2009} and a poor order choice can be very detrimental to the performance of the algorithm. In our case, the careful filtering procedure performed will eliminate all other multiple frequency resonances and additional spurious frequencies that may appear, leading to accurate estimates of the desired frequencies. 

The resonance extraction procedure, which we now describe, is summarized in Fig. \ref{fig:kalgorithm}.
We assume that the voltage output signal is sampled at a time-step that achieves equality in the Courant–Friedrichs–Lewy (CFL) stability condition of the FDTD algorithm~\cite{taf05}. That is, the maximum frequency component represented in the signal is $f_{\text{CFL}}/2$ where $f_{\text{CFL}}$ is the CFL frequency.  Also we assume that the excitation current pulse is given by (\ref{eq:gaussian}). A similar analysis can be carried out for other pulse shapes or starting sampling frequencies. In a first step, the output signal of the simulator is downsampled without an anti-aliasing filter. In general it is not necessary to use an antialiasing filter because the sampling frequency $f_{\text{CFL}}$ is much larger than the baseband bandwidth of the output signal, which is controlled by $f_p$. To find a reasonable downsampling factor, we note that the absolute value of the Fourier transform of $p(t)$ is:
\begin{equation}
	|P(f)| = \frac{1}{\sqrt{2f_p^2}} e^{-\frac{1}{2} \left(\frac{f}{f_p}\right)^2}.
\end{equation}
Thus, the amplitude of the input pulse normalized by its zero frequency value is:
\begin{equation}
	20 \log_{10} \left(\frac{|P(f)|}{|P(0)|}\right) = -10 \log_{10}(e) \left(\frac{f}{f_p}\right)^2  \ \ [\text{dB}]. \label{eq:attPulse}
\end{equation}
From (\ref{eq:attPulse}) it can be seen that the power content of the input signal above $5f_p$ will be more than 108dB below the zero frequency value. Thus, choosing a downsampling factor such that a new sampling frequency of $10f_p$ is achieved is a reasonable choice. In a second stage, the signal is downsampled again, this time using a low-pass filter, to a new sampling frequency of $4f_0$. After this downsampling stage, the maximum frequency present in the signal is therefore $2f_0$, that is, twice the resonance frequency of a single resonator. In a final filtering stage, a bandpass filter is applied to reduce noise and to further attenuate the resonance peaks which appear around $2f_0$.	In our case, we use a Gaussian bandpass filter, because it is straightforward to design in the time domain and to parameterize. Its impulse response is:
\begin{equation}
	h(t) = \cos\left(2\pi f_0 \left[t-\frac{1}{\tilde{f}}\right]\right) e^{-2 \pi \tilde{f}^2 \left(t-\frac{1}{\tilde{f}}\right)^2}, \label{eq:Gfilt}
\end{equation}
where $f_0$ is the resonance frequency of the resonators and $\tilde{f}$ is a parameter which controls the bandwidth of the filter. The requirement for $\tilde{f}$ is that it provide sufficient attenuation of the remains of the resonant peaks that are present near $2f_0$. In our examples we parameterize $\tilde{f}$ in terms of the bandwidth of the filter to be designed $B$ as:
\begin{equation}
	\tilde{f} = \alpha B,
\end{equation}
where $2 \leq \alpha \leq 5$ are reasonable values. This is because the coupling values required are related to the bandwidth of the filter. For stronger couplings, the frequencies $f_+$ and $f_-$ which appear around $f_0$ will be separated from $f_0$. This means that the double frequency resonances which appear around $2f_0$ will be closer to $f_0$ and will have a large chance of interfering with the estimation. For this reason, smaller values of $\alpha$ are more convenient for stronger couplings. On the contrary, for weak couplings, the peaks will be very close to $f_0$ and a larger value of $\alpha$ will be better to make sure that the bandpass filter does not suppress one of the peaks, if there is a small error in the estimation of the resonance frequency $f_0$. Finally, once the signal is properly filtered, the frequencies of the two resonant peaks are found by using a super-resolution algorithm. To estimate the frequency peaks we use ESPRIT based on the correlation matrix implementation using the forward-backward approach~\cite[Sec. 4.7]{stoica2005spectral}. In all cases we chose a correlation matrix of size 24 and extract 4 complex frequencies, which correspond to the two real frequency peaks that we want to estimate. As mentioned before, this can be done accurately thanks to the filtering procedure outlined before. Finally, the coupling coefficient is computed from the frequency estimates using (\ref{eq:K_eq}).

\begin{figure}
	\centering
	\begin{minipage}[]{0.45\columnwidth} \centering
		\includegraphics[width=1\linewidth]{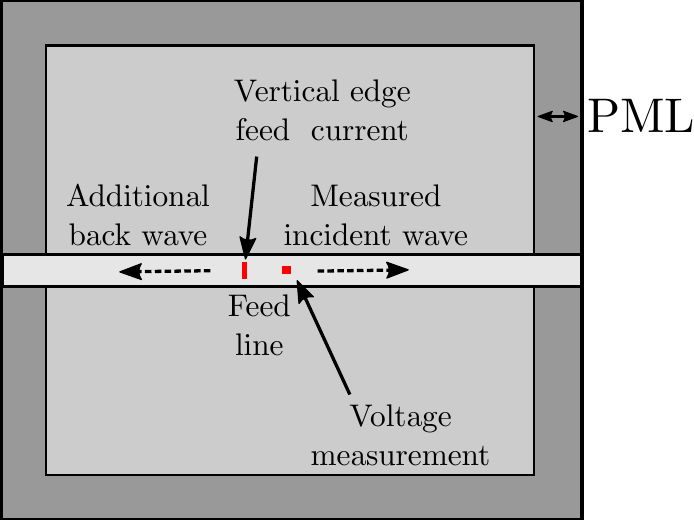}
		\subcaption{Setup to obtain the incident voltage. Top view of the substrate.}
	\end{minipage}
	\begin{minipage}[]{0.45\columnwidth}\centering
	\includegraphics[width=1\linewidth]{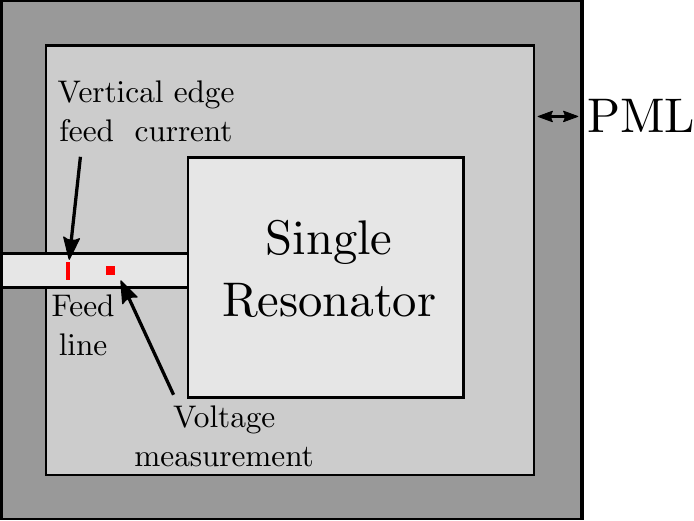}
	\subcaption{Setup for the obtain the total voltage. Top view of the substrate.}
	\end{minipage}
\caption{Setups to extract the external quality factor of a single resonator with its feed line.}
	\label{fig:qext}
\end{figure}

\subsection{Extraction of the External Quality Factor} \label{sec:Qe}
To extract the external quality factor it is more efficient to use the scattering parameters obtained from the FFT of the transient time signals. For completeness, we briefly review  the procedure here and discuss our implementation in gprMax. To extract the external coupling coefficient a single resonator is excited using the chosen feed method for the filter topology. Then, the $S_{1,1}$ parameter of this circuit is obtained and, from there, the external quality factor is obtained using the group delay as~\cite{Hong_microstripbook}:
\begin{equation}
	Q_{e,gd} = \frac{\omega_0 \tau_g(\omega_0)}{4}, \label{eq:Qe_gd}
\end{equation}
where $\omega_0$ is the resonance angular frequency of the resonator. Now, to extract the $S_{1,1}$ parameter we proceed in two stages. In the first stage, we extract an incident voltage  $V_{\text{inc}}$ by using the setup in Fig. \ref{fig:qext}a. In this setup, a short microstrip segment is excited in the middle using the feed model in Fig. \ref{fig:feedmodelperspective} which we used for the internal coupling coefficient. The microstrip and the substrate extend on both sides into the perfectly matched layer (PML) which means that the microstrip will be matched on both ends. When the current is injected, there will be an incident wave, whose voltage is measured, and a backwards wave, which is useless. Once the incident voltage is measured,  we implement the setup in Fig.  \ref{fig:qext}b. The resonator is connected to its feed line, and both the substrate and the feedline extend into the PML. The feed scheme is the same as before. When the current is injected, again there will be an incident wave which in this case moves towards the resonator and a backwards wave which vanishes in the PML. 
The voltage measurement $V_{\text{tot}}$ will include the superposition of the incident and reflected waves to and from the resonator, but will not include the backwards wave which moves toward the PML. Finally, the $S_{1,1}$ parameter is obtained from the FFT of the voltage waves as\cite{Kong1990}:
\begin{equation}
	S_{1,1} (f) = \frac{ \mathfrak{F} (V_{\text{tot}}-V_{\text{inc}})}{\mathfrak{F} (V_{\text{inc}})},
\end{equation}
where $ \mathfrak{F}$ is the Fourier Transform of the time signals.

\section{Simulation and Measurement Comparisons} \label{sec:example}

\subsection{General remarks and simulation setup}
\begin{figure}[t!]
	\begin{minipage}[b]{.48\linewidth}
		\centering
		\includegraphics[width=0.7\linewidth]{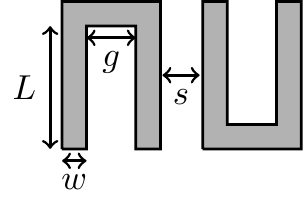}
		\subcaption{Size parameterization of two resonators.}
		\label{fig:hairpinsize}
	\end{minipage}
	\begin{minipage}[b]{.5\linewidth}
		\centering
		\includegraphics[width=0.7\linewidth]{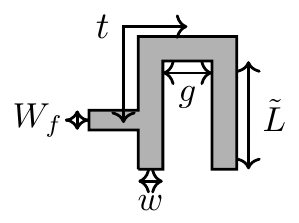}
		\subcaption{Size parameterization of a resonator with a feed line. }
		\label{fig:hairpinsizeQ}
	\end{minipage}
	\caption{Parameterization of the hairpin resonators for simulations. }
	\label{fig:hairpin_size} \vspace{-3mm}
\end{figure}

In this section we study the performance the proposed approach using time domain simulations.  In Section \ref{sec:comparison} we compare our results for high and low couplings to simulations presented in other papers. Then in Section \ref{sec:measured} we design and build a filter to validate the procedure using real measurements. All the examples are presented for hairpin filters which are one of the most representative examples, but other resonators could also be considered with similar results. For both sections we parameterize the shape of the hairpin resonators according to Fig. \ref{fig:hairpin_size}.

\begin{figure}
	\centering
	\includegraphics[width=.8\linewidth]{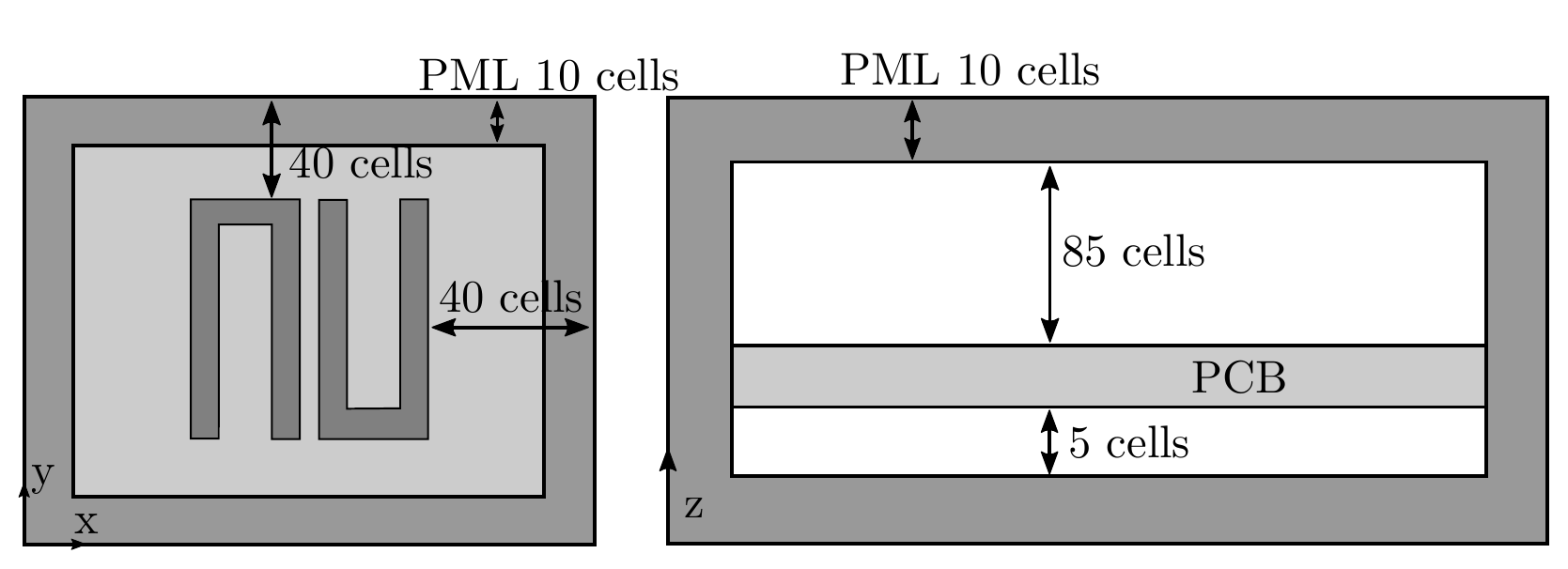}
	\caption{Example layout for extraction of type 1 mixed coupling.}
	\label{fig:mixed coupling layout}
\end{figure}

As mentioned before, all our FDTD simulations were carried out in gprMax~\cite{WARREN2016163} v3.1.5 using the GPU engine~\cite{WARREN2019208} which allows for substantial improvement in computation speed. In all the simulations, the $z$ direction of the coordinates is assumed to be normal to the PCB plane, which is aligned with the $x/y$ plane. The simulation domain is always terminated in a PML with a thickness of 10 cells using the standard PML formulation of gprMax~\cite{GiannoPML}. In all the simulations to extract the coupling coefficients, there is a clearance of 40 cells in every direction of the $x/y$ plane, and a vertical clearance above the PCB of 95 cells, including the PML (see Fig. \ref{fig:mixed coupling layout} for an example). 

For the extraction of the external quality factor, the procedure outlined in Section \ref{sec:Qe} is used (see Fig. \ref{fig:qext} for the layout). The feed line length is always 50 cells (including the 10 cells of the PML), the current source is located 5 cells away from the PML and the voltage is measures 15 cells away from the PML (10 from the current source). The vertical and additional horizontal clearances are the same as for the extraction of the coupling coefficients. 

All the simulations were done using the dielectric smoothing option enabled, which is fundamental for accurate results. All microstrips are modeled as zero-thickness perfect electric conductor (PEC) strips except for the measured filter, where copper is used.


\subsection{Example of comparison to other works} \label{sec:comparison}
\begin{figure}[t!]
	\centering{
		\begin{minipage}[]{0.45\columnwidth} \centering
			\includegraphics[width=0.6\linewidth]{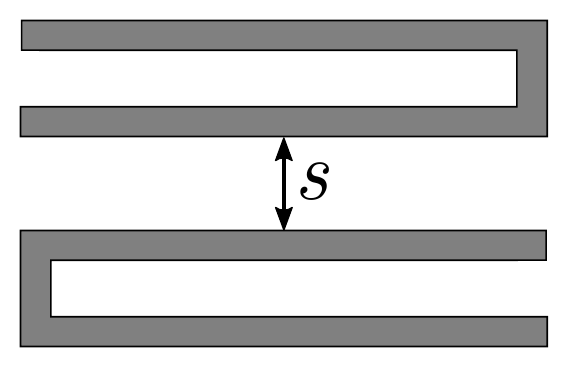}
			\subcaption{First type of mixed coupling.}
		\end{minipage}
		\begin{minipage}[]{0.45\columnwidth} \centering
			\includegraphics[width=0.6\linewidth]{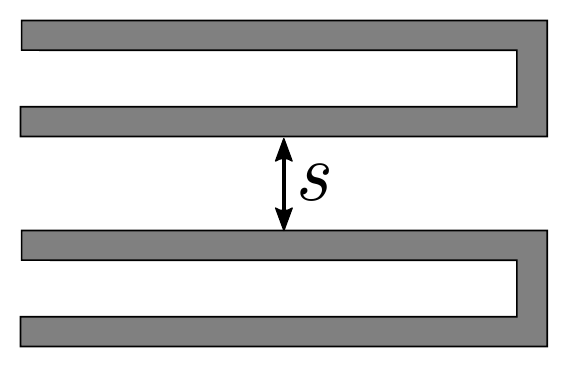}
			\subcaption{Second type of mixed coupling.}
		\end{minipage}
		\includegraphics[width=0.6\linewidth]{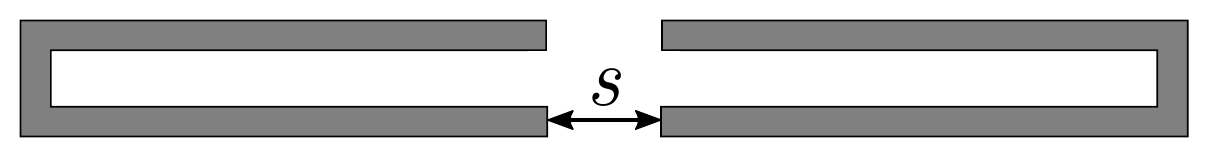}
		\subcaption{Electric coupling.}
		
		\includegraphics[width=0.6\linewidth]{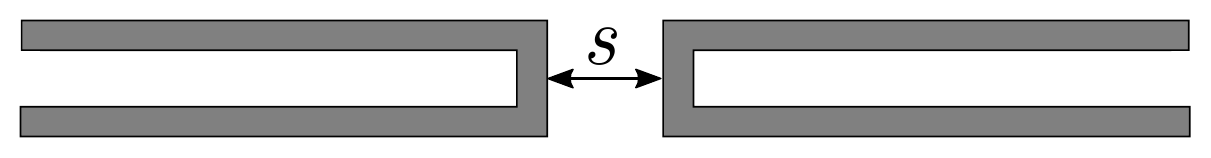}
		\subcaption{Magnetic coupling.}
		
	}
	\caption{Different type of couplings for microstrip hairpin resonators, depending on the separation $s$.}
	\label{fig:Kmodes}
\end{figure}

\begin{figure}[t!]
	\centering
	{\begin{tikzpicture}
			\begin{axis}[scale=1,xlabel={$s$ [mm]}, ylabel={$K$},height=4cm,width=9cm,
				legend pos = north east,
				ymin =0.0, ymax = 0.1,/pgf/number format/precision=3,
				xmin=1,xmax=3,
				minor x tick num=1,minor y tick num=3, grid =both,
				legend entries = {{\cite{Hong98} Fig. 3},Proposed TD approach}]%
				\addplot[red,line width=1pt]  table {Dats/Hong1G/final/Kmixed1_hong_rescaled.txt} ;
				\addplot[blue, mark=o,mark size =2pt,line width=1pt, only marks]  table {Dats/Hong1G/final/Kmixed1_points.txt} ;
				\addplot[blue,mark size =1.5pt,line width=1pt]  table {Dats/Hong1G/final/Kmixed1_interp.txt} ;
				

			\end{axis}
		\end{tikzpicture}
		\subcaption{First type of mixed coupling (Fig. \ref{fig:Kmodes}(a)) as a function of the separation $s$.}
		
		\begin{tikzpicture}
			\begin{axis}[scale=1,xlabel={$s$ [mm]}, ylabel={$K$},height=4cm,width=9cm,
				legend pos = south east,
				ymin =0.0, ymax = 0.02,/pgf/number format/precision=3,
				xmin=0.5,xmax=3,
				minor x tick num=1,minor y tick num=3, grid =both,
				legend entries = {{\cite{Hong98} Fig. 3},Proposed TD approach}]
				\addplot[red,line width=1pt]  table {Dats/Hong1G/final/Kmixed2_hong_rescaled.txt} ;
				\addplot[blue, mark=o,mark size =2pt,line width=1pt, only marks]  table {Dats/Hong1G/final/Kmixed2_points.txt} ;	
				\addplot[blue,mark size =1.5pt,line width=1pt]  table {Dats/Hong1G/final/Kmixed2_interp.txt} ;

				
			\end{axis}
		\end{tikzpicture}
		\subcaption{Second type of mixed coupling (Fig. \ref{fig:Kmodes}(b)) as a function of the separation $s$.}

		\begin{tikzpicture}
			\begin{axis}[scale=1,xlabel={$s$ [mm]}, ylabel={$K$},height=4cm,width=9cm,
				legend pos = north east,
				ymin =0.0, ymax = 0.03,
				xmin=0.25,xmax=3,/pgf/number format/precision=3,
				minor x tick num=1,minor y tick num=3, grid =both,
				legend entries = {{\cite{Hong98} Fig. 3},Proposed TD approach}]
				\addplot[red,line width=1pt]  table {Dats/Hong1G/final/Kelectric_hong_rescaled.txt} ;
				\addplot[blue, mark=o,mark size =2pt,line width=1pt, only marks]  table {Dats/Hong1G/final/Kelect_points.txt} ;
				\addplot[blue,mark size =1.5pt,line width=1pt]  table {Dats/Hong1G/final/Kelect_interp.txt} ;
			\end{axis}
		\end{tikzpicture}
		\subcaption{Electrical coupling (Fig. \ref{fig:Kmodes}(c)) as a function of the separation $s$. }	
		\begin{tikzpicture}
			\begin{axis}[scale=1,xlabel={$s$ [mm]}, ylabel={$K$},height=4cm,width=9cm,
				legend pos = north east,
				ymin =0.0, ymax = 0.02,
				xmin=0.25,xmax=3,
				minor x tick num=1,minor y tick num=3, grid =both,
				legend entries = {{\cite{Hong98} Fig. 3,Proposed TD approach}}]
				\addplot[red,line width=1pt]  table {Dats/Hong1G/final/Kmagnetic_hong_rescaled.txt} ;
				\addplot[blue, mark=o,mark size =2pt,line width=1pt, only marks]  table {Dats/Hong1G/final/Kmagnet_points.txt} ;
				\addplot[blue,mark size =1.5pt,line width=1pt]  table {Dats/Hong1G/final/Kmagnet_interp.txt} ;
			\end{axis}
		\end{tikzpicture}
		\subcaption{Magnetic coupling coupling factor  (Fig. \ref{fig:Kmodes}(d) as a function of the separation $s$.}
	}
	\caption{Example of extraction of electrical coupling between hairpin resonators. Substrate with $\epsilon_r = 10.8$, $h=1.27$mm. $L=27.5$ mm, $w=1.5$mm, $g=2.5$mm. Reference curves are taken from \cite{Hong98} Fig. 3, obtained using the method of moments.}
	\label{fig:KxHong1G}\end{figure}
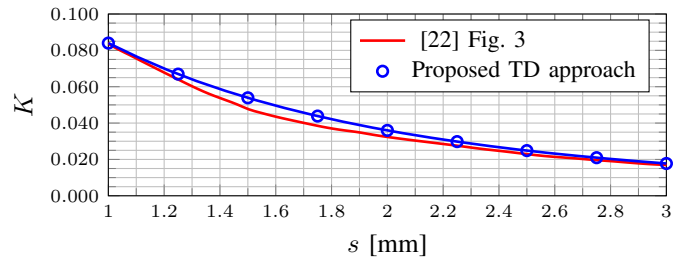
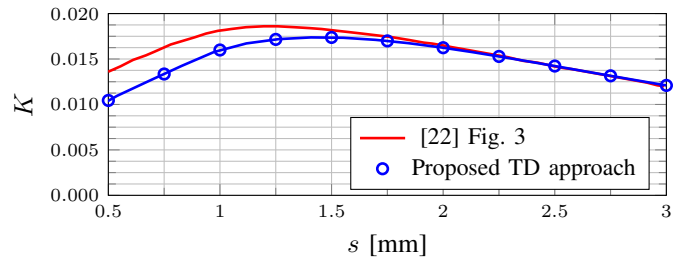
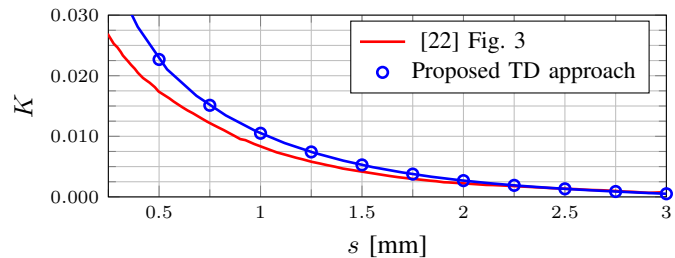
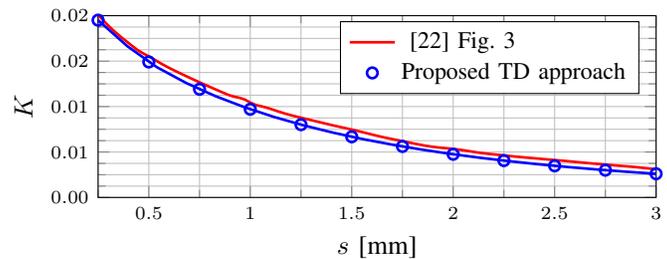

We now apply the proposed procedure to extract the four different types of couplings of hairpin resonators which can be seen in Fig. \ref{fig:Kmodes}. All of the coupling modes will become increasingly harder to extract as they become weaker, since the spectral peaks will become very close and will be harder to separate.
We compare our results with the ones presented in \cite{Hong98}, Fig. 3. The goal there is to design a cross-coupled hairpin filter with central frequency $f_c=965$MHz and a bandwidth $B=20$MHz. The hairpin parameters are $L=27.5$mm, $g=2.5$mm and $w=1.5$mm, and a substrate of height $h=1.27$mm and dielectric constant $\epsilon_r = 10.8$ is used.
For our simulations, the FDTD cell sizes were chosen to be $\Delta x = \Delta y = 0.125$mm and $\Delta_z = 0.127$mm. Thus, the substrate height will fit in an integer number of cells. The simulations are run for a total of 100ns (413236 time steps) and the resonators are excited using a Gaussian pulse with $f_p=5$GHz.  To extract the coupling coefficients, we use the procedure outlined in Section \ref{sec:extractionK}. We use a Gaussian bandpass filter given by (\ref{eq:Gfilt}) with $\tilde{f} = 5B$, which from (\ref{eq:attPulse}) gives an attenuation more than 400dB at $2f_c$. In Fig. \ref{fig:KxHong1G} we can see the comparison of our approach with the simulations from Fig. 3 in \cite{Hong98}, which were obtained with the method of moments.
We can see that the proposed approach provides a very good fit to previously published results in all the cases. For a prescribed coupling coefficient, the difference in separation provided by each approach is very similar. The largest difference appears in the case of type 2 mixed couplings, where the worst difference is around 0.25mm for a coupling of $K\approx 0.014$. 

\subsection{Measured filter} \label{sec:measured}
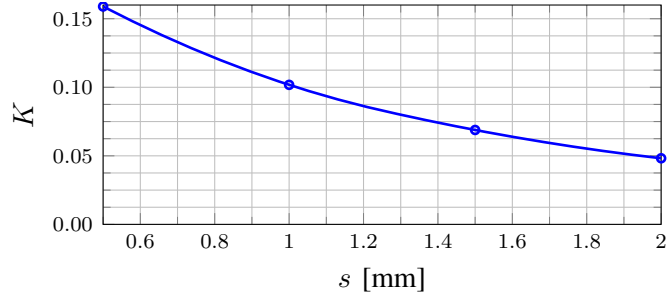
\begin{figure}[t!]
	\centering
	\begin{tikzpicture}
		\begin{axis}[scale=1,xlabel={$s$ [mm]}, ylabel={$K$},height=4.5cm,width=9cm,
			legend pos = north east,
			ymin =0.00, ymax = 0.16,
			xmin=0.5,xmax=2,
			minor x tick num=1,minor y tick num=3, grid =both,
			]
			
			\addplot[blue,mark size =1.5pt,line width=1pt]  table {Dats/Measured/Kmeasured_interp40.txt} ;
			\addplot[blue, mark=o,mark size =1.5pt,line width=1pt, only marks]  table {Dats/Measured/Kmeasured_raw40.txt} ;
			
		\end{axis}
	\end{tikzpicture}
	\caption{Internal coupling factor $K$ as a function of separation distance. Substrate with $\epsilon_r = 3.52$, $h=1.524$mm. $L=9.9$mm, $w=1.5$mm, $g=3$mm.}
	\label{fig:Kx_example}
\end{figure}

In this section we build a filter using the proposed approach and we compare the simulation of the built filter with the measurements. We chose a 5th order Butterworth filter of center frequency $f_c=3.75$GHz, a  bandwidth $B=350$MHz, and with 50$\Omega$ tapped-line feed \cite{Wong79}. The layout for this type of filter and the couplings to be found can be seen in Fig. \ref{fig:hairpin1}. For this design, a Rogers RO4350 substrate with a  dielectric constant $\epsilon_r=3.48$, thickness $h=1.524$mm is chosen. 
The normalized elements of the desired response are:
\begin{center}
	\begin{tabular}{c|c|c|c}
		$g_0, g_6$ & $g_1,g_5$ & $g_2, g_4$ & $g_3$ \\\hline \hline
		1 & 0.618 & 1.618& 2
	\end{tabular}.
\end{center}
These values are converted to the required couplings via the following equations~\cite{MATTHAEI}:
\begin{gather}
	Q_{e,1} = \frac{g_0 g_1}{\Delta}, \ \ \ Q_{e,n} =  \frac{g_n g_{n+1}}{\Delta}, \label{eq:Qext}\\
	K_{i, i+1} = \frac{\Delta}{\sqrt{g_i g_{i+1}}} \ \ \ i = 1,\ldots,n-1. \label{eq:K}
\end{gather}
This leads to the following required coupling coefficients:
\begin{center}
	\begin{tabular}{c|c|c}
		$Q_{e,1}, Q_{e,5}$ & $K_{1,2}, K_{3,4} $ & $K_{2,3}, K_{4,5}$ \\\hline \hline
		6.62 & 0.093 &0.052
	\end{tabular}
\end{center}
The width of the resonator is chosen to be $w=1.5$mm (characteristic impedance around 75$\Omega$) and $g=2w$. The resonator length 
$L$  was chosen to be $L=9.9$mm. 
For all the simulations the cell size are chosen as $\Delta x = \Delta y = 0.1$mm, $\Delta z=0.1524$mm, and a Gaussian pulse with $f_p = 10$GHz is used. To extract the coupling coefficient the Gaussian bandpass filter is chosen to have $\tilde{f}=2B$, to minimize the effect of the double frequency resonances which are closer to $f_c$ in this case.
The coupling coefficient as a function of separation is then obtained by running each configuration for a total of 10ns. The obtained results can be seen in Fig. \ref{fig:Kx_example}.  

To extract the external quality factor, we simulated the feed line and the first resonator from Fig. \ref{fig:hairpin1} using the procedure indicated in Section \ref{sec:Qe}. Each simulation was run for 10ns, and we extracted the external quality factor as a function of the tapping point $t$. The results can be seen in Fig. \ref{fig:Qe_example}.

\begin{figure}[t!]
	\centering
	\begin{tikzpicture}
		\begin{axis}[scale=1,xlabel={$t$ [mm]}, ylabel={$Q_e$},height=4.5cm,width=9cm,
			legend pos = north east,
			ymin =0, ymax = 7.5,
			xmin=4.75,xmax=10,
			minor x tick num=1,minor y tick num=3, grid =both,
			legend style = {font = \scriptsize},]
			\addplot[blue,mark=o,mark size =1.5pt,line width=1pt]  table {Dats/Measured/Qext.txt} ;
		\end{axis}
	\end{tikzpicture}
	\caption{External coupling factor $Q_e$ as a function of the tapping point $t$ from (\ref{eq:Qe_gd}) using the group delay. Substrate with $\epsilon_r = 3.66$, $h=1.524$mm. $L=9.9$mm, $w=1.5$mm, $g=3$mm.}
	\label{fig:Qe_example} \vspace{-3mm}
\end{figure}
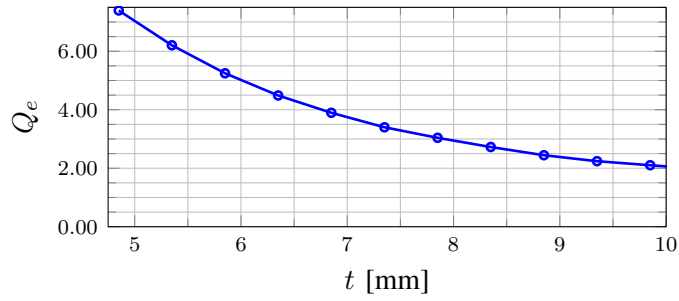
\begin{table}[t!]
	\begin{center}
		\begin{tabular}{c|c|c|c|c|c|c|c}
			&$w$ & $g$  &$L$& $s_1$ & $s_2$ & $t$& $\tilde{L}$ \\\hline \hline
			Design value & 1.5 & 3 & 9.9& 1.1 & 1.9 & 5.35  & 9.9 \\
			Final value & 1.5 & 3 & 9.9 & 1.1 & 1.9 & 5.35 & 9.4
		\end{tabular}
	\end{center}
	\caption{Initial and final filter dimensions for the  filter designed in Section \ref{sec:measured}.}
	\label{tab:f_values}
\end{table}
Using Figs. \ref{fig:Kx_example} and \ref{fig:Qe_example} the hairpin separation and tapping points were selected.  
In order to simulate the filters, we needed to set the loss model for the substrate. Since we cannot choose a constant loss tangent model for gprMax, we used a second order Debye model which we tuned so that the dielectric constant and loss tangent were close to the values of the substrate in the prescribed frequency range. The model for the  permittivity used is therefore:
\begin{equation}
	\epsilon (\omega) = \epsilon_{\inf} + \sum_{m=1}^2 \frac{\Delta \epsilon_i}{1 + j\omega \tau_m},
\end{equation} 
where $\tau_1= 9.8\times 10^{-11}s$, $\tau_2 = 1.56\times 10 ^{-11}s$, $\Delta \epsilon_1= \Delta\epsilon_2 = 0.0164$, $\epsilon_{\inf} = 3.5$.
Using this loss model, the value of $\tilde{L}$ was adjusted by numerical simulations to optimize the response of the filter, starting from the initial value $\tilde{L} = L$. The feed model used for the filter was the same as for the extraction of the external quality factor. The feed line length and horizontal clearance around the filter was 100 cells (including the PML) and the position of the current source and voltage sensor was the same as for the external quality factor. The initial and final values after this optimization can be seen in Table \ref{tab:f_values}. As can be seen, the only difference between the initial parameters is that the first and last hairpins were shortened by 0.5mm. A picture of the manufactured filter can be seen in Fig. \ref{fig:filtrocrop2}. Finally, the comparison between the simulated and measured values can be seen in Fig. \ref{fig:filt3800}. It can be seen that the simulation and measurements are in good agreement. The simulated filter has a passband loss of 1.1dB, while the measured filter has an additional 0.6dB of loss. 
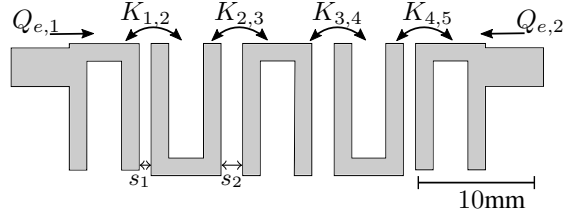
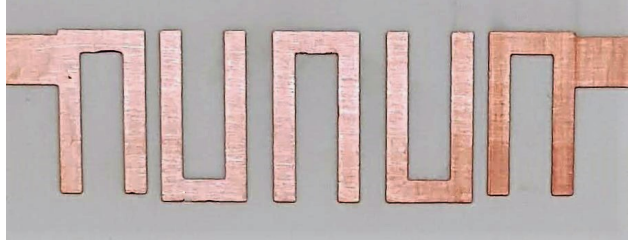
\begin{figure}[t!]
	\centering {
	\begin{tikzpicture}
				\node at (0,0) {\includegraphics[width=0.8\columnwidth]{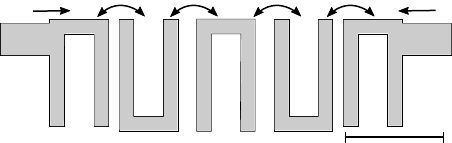}};
				\node at(-3.2, 1.1) {$Q_{e,1}$};
				\node at(-1.75, 1.2) {$K_{1,2}$};
				\node at(-0.5, 1.2) {$K_{2,3}$};
				\node at(0.8, 1.2) {$K_{3,4}$};
				\node at(2, 1.2) {$K_{4,5}$};
				\node at(3.5, 1.1) {$Q_{e,2}$};
				\node at(2.85, -1.25) {$10$mm};
				\draw[<->] (-1.83,-.8)  node[below,font=\small] {$s_1$}-- (-1.68,-.8);
				\draw[<->] (-.74,-.8)  -- node[below, font=\small] {$s_2$}(-.47,-.8);
			\end{tikzpicture}
		\subcaption{Layout of a symmetrical 5th order bandpass filter with tapped-line feed.} \label{fig:hairpin1}
\includegraphics[width=0.5\linewidth]{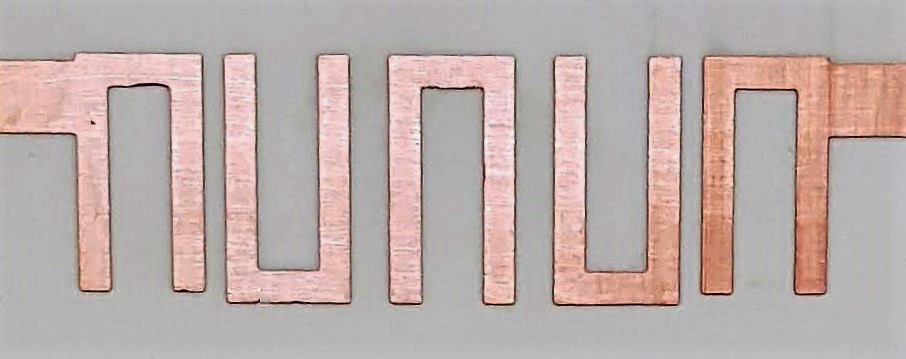}
\subcaption{Manufactured linear hairpin filter on a RO4350 substrate.}
\label{fig:filtrocrop2}
}
\caption{Layout for the designed filter and photograph of the constructed prototype. The filter parameterization can be seen in Fig. \ref{fig:hairpin_size} and its relevant dimensions in Table \ref{tab:f_values}.}
\end{figure}

\begin{figure}[t!]
	\centering
	\begin{tikzpicture}
		\begin{axis}[scale=1,xlabel={$f$ [GHz]}, ylabel={ [dB]},height=5cm,width=9cm,legend pos = south west,
			ymin =-56, ymax = 1,xmin=3,xmax=4.5,legend columns=2,	yticklabel style={/pgf/number format/precision=0},
			minor x tick num=1,minor y tick num=3, grid =both,
			legend entries = {{$S_{1,1}$  FDTD },{$S_{2,1}$ FDTD},{$S_{1,1}$ Measured},{$S_{2,1}$ Measured}},
			legend style = {font = \scriptsize,at={(0.005,0.005)}},]
			\addplot[blue, mark size =1pt,,line width=1pt]  table[x index=0, y index=1] {Dats/Measured/Spars_final_sim.txt} ;
			\addplot[black,dashed,mark size =1pt,,line width=1pt]  table[x index=0, y index=2] {Dats/Measured/Spars_final_sim.txt} ;

			\addplot[orange,mark=o,mark repeat=60,mark size =1pt,,line width=1pt]  table[x expr=\thisrowno{0}/1e9, y index=1] {Dats/filter_measured.txt} ;
			\addplot[red,mark=x,mark repeat=60,mark size =1.5pt,,line width=1pt]  table[x expr=\thisrowno{0}/1e9, y index=2] {Dats/filter_measured.txt} ;		
		\end{axis}
	\end{tikzpicture}
	\caption{Measured filter, compared to final filter design after tuning of $\tilde{L}$. } \vspace{-3mm}
	\label{fig:filt3800}
\end{figure}
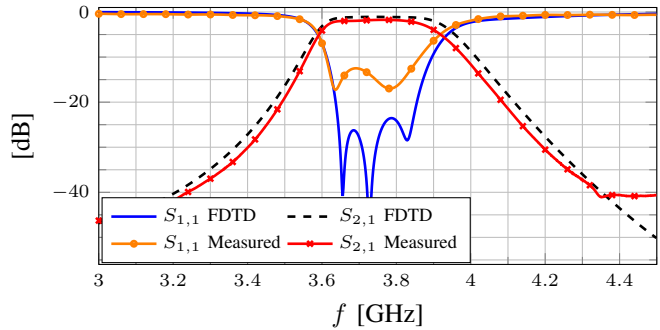
\section{Conclusions}
In this journal we have presented a procedure to extract the coupling coefficient between microstrip resonators for the design of filters by using time domain simulations and super-resolution techniques. We have shown through examples that this technique produces results which are very similar to those obtained through frequency domain approaches. This was shown to be true even in the case of very low couplings, where a naive FFT approach from time domain signals would require extremely long and unrealistic simulation times. This procedure is simple, requiring a minimal adjustment of parameters, and provides a substantial reduction in simulation times compared to the FFT approach. Also we have shown that it provides a good performance compared to measurements.

\bibliographystyle{IEEEtran}
\bibliography{IEEEabrv,biblio3}
\end{document}